\documentclass[11pt]{amsart}
\usepackage{geometry}                
\usepackage{graphicx}
\usepackage{amssymb}
\title{An empirical partition function for the simple cubic Ising Model with a zero external magnetic field}
\author{Rong Qiang Wei}
\address{College of Earth and Planetary Sciences, University of Chinese Academy of Sciences, Beijing, PRC, 100049}
\email{wrq1973@ucas.edu.cn}
\date{}

\begin{document}
\maketitle

\begin{abstract}
  There is no an accepted exact partition function (PF) for the three dimensional (3D) Ising model to our knowledge. Mainly based on the connection between the lattice Green function (LGF) for the simple cubic lattice and that for the honeycomb lattice, we infer an empirical partition function (EPF)  for the simple cubic Ising model in the absence of an external magnetic field. This ${\rm EPF}_{_{\rm 3D}}=\frac{1}{{2{\pi ^3}}}\int_0^\pi  \int_0^\pi  \int_0^\pi  \log [2(2{{\cosh }^3}2z + 3{{\sinh }^2}2z + 2)^{\frac{1}{2}} -2\alpha\sinh 2z \  (\cos\omega_1 + \cos \omega_2 + \cos\omega_3)]  {\rm{d}}\omega_1{\rm{d}}\omega_2{\rm{d}}\omega_3, \alpha\in[\sqrt{2},\sqrt{3}]$ (where $z=\frac{\epsilon}{kT}$, $\epsilon$ the interaction energy, $T$ the temperature, and $k$ Boltzmann constant). When $\alpha=\sqrt{2}$, this EPF is consistent well numerically with the result from high temperature expansions by Guttmann and Enting (1993). The specific heat from this EPF approaches infinity non-logarithmically at the critical temperature $T_c$. $\frac{\epsilon}{kT_c}=\cosh^{-1}[\frac{1}{4} (17-3\sqrt{17})]/2\approx 0.277212$, which is greater than 0.221654 from the recent Monte Carlo study.
  
\end{abstract}

{\hspace{2.2em}\small Keywords:}

{\hspace{2.2em}\tiny 3D Ising model, Empirical partition function, High temperature expansions, Lattice Green function}

\section{Introduction}

The Ising model is an important statistic mechanical model for simulating the ferromagnetic system. This model consists of a lattice with a binary magnetic polarity (or "spin") assigned to each point.  The nearest-neighbor Ising model without an external magnetic field in $D$-dimensions ($D=1,2,3,...$) is defined in terms of the following Hamiltonian (eg., Huang, 1987),

\begin{equation}\label{eq1}
\mathcal{H} =  -\frac{1}{2}\sum\limits_{i,j = 1}^N {{K_{ij}}} {s_i}{s_j}
\end{equation}
where, $i$ and $j$ are the sites $\bf{r}_i$ and $\bf{r}_j$ of a $D$-dimensional hyper cubic lattice with $N$ sites, respectively. $s_i=\pm 1$ are the two possible states of the $z$-components of spins localized at the lattice sites. $K_{ij}$ denotes the exchange interaction between spins localized at $\bf{r}_i$ and $\bf{r}_j$, 

\begin{equation}\label{eq2}
{K_{ij}} = \left\{ {\begin{array}{ll}
z&{{\rm{if \hspace{0.5em}}}i{\rm{\hspace{0.5em}and\hspace{0.5em}}}j{\rm{\hspace{0.5em} are\hspace{0.2em} the \hspace{0.2em}nearest\hspace{0.2em} neighbors}}}\\
0&{{\rm{otherwise}}}
\end{array}} \right.
\end{equation}
where $z=\beta\epsilon=\frac{\epsilon}{kT}$, $\epsilon$ the interaction energy, $T$ the temperature, and $k$ Boltzmann constant.

Such a model has played a special role in the theory of ferromagnetism and phase transitions which depends on the evaluation of the partition function. For the one dimensional (1D) Ising model with periodic boundary condition, the exact PF, $\frac{1}{N}\log Q(0,T)$, can be obtained from the solution of Kramers and Wannier (1941),

\begin{equation}\label{KW}
\frac{1}{N}\log Q_{_{1d}}(0,T)=\log\left[2\cosh z \right]
\end{equation}

For the two dimensional (2D) Ising model imposed on periodic boundary condition, the exact PF was evaluated by Onsager (1944) as the following, 

\begin{equation}\label{Onsager}
\frac{1}{N}\log Q_{_{2d}}(0,T)=\log(2\cosh 2z)+\frac{1}{2\pi}\int_0^\pi {\rm{d}}\phi\log\frac{1}{2}\left(1+\sqrt{1-\kappa^2\sin^2\phi}\right)
\end{equation}
where $$\kappa=\frac{2}{\cosh 2z \coth 2z}$$

However,  there is still no an accepted exact solution for the 3D Ising model especially the one which is both elegant and exact like Eq. (\ref{KW}) or (\ref{Onsager}), although there have been many efforts. For example, Zhang (2007) published an exact PF that he called “conjecture”, but many physicists think this conjecture is incorrect (eg., Wu et al., 2008a; Wu et al., 2008b; Fisher and Perk, 2016). Kocharovsky and Kocharovsky (2015) found the consistency equations for the main steps in the analysis of the 3D Ising model, and they said that "Towards an exact solution for the three-dimensional Ising model". However, "their exact results are for very small lattices and to go to larger lattices requires exponentially growing effort"\footnote{\footnotesize Private e-mail communication with Perk, J. H..}. By using the Hubbard-Stratonovich transformation, Wei (2018) presented a simple but exact solution to the PF for the finite-size 3D Ising model. However, this solution looks formidable and unintuitive because it is expressed in a sum of $2^N$ exponential functions (See Eq. (\ref{wei3d}) below). For a large $N$, it is impracticable to calculate the PF.  What's even more frustrating is that Cipra (2000) has even claimed that the exact solution for the 3D Ising model is essentially an NP-complete problem. If so, this means that we can not expect a simple closed-form solution like Eq. (\ref{KW}) or (\ref{Onsager}).

In the absence of a practicable exact solution, an EPF, which based on reasonable facts and is under some reliable constraints, is one of the possible ways to understand the 3D Ising model. Here we infer such an EPF for the Ising model on a simple cubic lattice without an external magnetic field, and investigate its critical temperature and the singularity at this temperature.

\section{The form of the EPF for the simple cubic Ising Model}\label{sec2}

We infer that the EPF for the simple cubic Ising Model has the following form,

 \begin{equation}\label{Eq_infer}
\frac{1}{N}\log Q_{_{3d}}(0,T)=\frac{1}{{2{\pi ^3}}}\int_0^\pi  {\int_0^\pi  {\int_0^\pi  {\log [2f(z)  - 2g(z)(\cos\omega_1 + \cos \omega_2 + \cos\omega_3)]} } } {\rm{d}}\omega_1{\rm{d}}\omega_2{\rm{d}}\omega_3
\end{equation}
where  $f(z)$ and $g(z)$ are the functions to be inferred further.

It should be pointed out firstly we call Eq. (\ref{Eq_infer}) EPF because it is not obtained by rigorous derivation but empirical inferring, and there are other empirical PFs for the simple cubic Ising model.  

The reasons are as follows,

\subsection{Eq. (\ref{KW}) or (\ref{Onsager}) can be transformed into the form of Eq. (\ref{Eq_infer})}

\ \ \ 

a.  For 1D Ising model,

 \begin{equation}\label{KW1}
\begin{array}{ll}
 \frac{1}{N}\log Q_{_{1d}}(0,T) &={\log (2\cosh z)}\\
{}&{}\\
  &={\frac{1}{2}\log (2\sinh z) + \frac{1}{2}\ln (2\coth z\cosh z)}\\
{}&{}\\
  &\approx{\frac{1}{2}\log (2\sinh z) + \frac{1}{2}{{\cosh }^{ - 1}}(\coth z\cosh z)}\\
{}&{}\\
  &={\frac{1}{2}\log (2\sinh z) + \frac{1}{{2\pi }}\int_0^\pi  {\log (2\coth z\cosh z - 2\cos {\omega _1}){\rm{d}}{\omega _1}} }\\
{}&{}\\
  &={\frac{1}{{2\pi }}\int_0^\pi  {\log (2\cdot 2{{\cosh }^2}z - 2\cdot 2\sinh z\cos {\omega _1}){\rm{d}}{\omega _1}} }\end{array}
\end{equation}
where $\cosh^{-1}z=\frac{1}{\pi}\int_0^\pi \log(2z-2\cos\omega_1){\rm{d}}\omega_1$ and $\cosh^{-1}z\approx \log 2z \ \ (z \ge 1)$ are used.

\ \ \ 

b. For 2D Ising model, according to Huang (1987) and Martin (1991), 

 \begin{equation}\label{Onsager1}
\begin{array}{ll}
\frac{1}{N}\log Q_{_{2d}}(0,T)&{ = \frac{1}{2}\log (2\sinh 2z) + \frac{1}{{2{\pi ^2}}}\int_0^\pi  {\int_0^\pi  {\log (2\cosh 2z\coth 2z - 2\cos {\omega _1} - 2\cos {\omega _2}){\rm{d}}{\omega _1}{\rm{d}}{\omega _2}} } }\\
{}&{}\\
{}&{ = \frac{1}{{2{\pi ^2}}}\int_0^\pi  {\int_0^\pi  {\log [2\cdot 2{{\cosh }^2}2z - 2\cdot 2\sinh 2z{\rm{ }}(\cos {\omega _1} + \cos {\omega _2})]{\rm{d}}{\omega _1}{\rm{d}}{\omega _2}} } }
\end{array}
\end{equation}

\subsection{Eigenvalues of matrix {\bf K} in the Ising Hamiltonian}

\ \ \ \ 

\ \ \ \ 

From Wei (2018), we know the PF for a $N$-sites Ising model without an external magnetic field is, 

\begin{equation}\label{wei3d}
\begin{array}{ll}
Q(0,T) = & \exp \left( {\frac{1}{2}{\bf{K}}_0{{\bf{K}}^{ - 1}}{{\bf{K}}_0^T}} \right) +
\sum\limits_{\alpha  = 1}^N \exp \left[ {\frac{1}{2}({\bf{K}}_0-2{\bf{K_\alpha}}){{\bf{K}}^{ - 1}}{({\bf{K}}_0-2{\bf{K_\alpha}})^T}} \right]+\\
& \sum\limits_{\begin{array}{cc}
{\tiny\alpha ,\beta  = 1}\\
{\alpha  < \beta }
\end{array}}^N \exp \left[ {\frac{1}{2}({\bf{K}}_0-2{\bf{K_\alpha}}-2{\bf{K_\beta}}){{\bf{K}}^{ - 1}}{({\bf{K}}_0-2{\bf{K_\alpha}}-2{\bf{K_\beta}})^T}} \right]+\\
& \sum\limits_{\begin{array}{cc}
{\alpha ,\beta ,\gamma  = 1}\\
{\alpha  < \beta  < \gamma }
\end{array}}^N \exp \left[ {\frac{1}{2}({\bf{K}}_0-2{\bf{K_\alpha}}-2{\bf{K_\beta}}-2{\bf{K_\gamma}}){{\bf{K}}^{ - 1}}{({\bf{K}}_0-2{\bf{K_\alpha}}-2{\bf{K_\beta}}-2{\bf{K_\gamma}})^T}} \right]+
\ldots \\
& + \exp \left[ {\frac{1}{2}(-{\bf{K}}_0){{\bf{K}}^{ - 1}}{(-{\bf{K}}_0)^T}} \right]
 \end{array}
\end{equation}
where ${\bf{K}}_0=\sum\limits_i^N {\bf K}_i$ 

Eq. (\ref{wei3d}) reads that the PF for the Ising model is only dependent on the matrix {\bf K} in the Ising Hamiltonian in Eq. (\ref{eq1}). 

When {\bf K=A}, Eq. (\ref{wei3d}) is the PF of the 1D Ising model, and {\bf A} is an $N\times N$  matrix (eg., Dixon et al., 2001), 

\[{\bf A} =z\times \left[ {\begin{array}{*{20}{c}}
0&1&0&0& \cdot & \cdot & \cdot &1\\
1&0&1&0&{}&{}&{}&{}\\
0&1&0&1&{}&{}&{}&{}\\
0&0&1&0&{}&{}&{}&\cdot\\
 \cdot &\cdot&\cdot&{}&{}&{}&{}&\cdot\\
 \cdot &\cdot&\cdot&{}&{}&{}&{}&\cdot\\
 \cdot &\cdot&\cdot&{}&{}&{}&{}&1\\
1&0&0&{}&{}&{}&1&0
\end{array}} \right]\]

with the $r$-th ($r=1,2,..., N$) eigenvalue is (eg., Berlin and Kac, 1952),

\[{E_{r}} = 2z\left\{ {\cos \left[ {\frac{{2\pi }}{N}(r - 1)} \right]} \right\}\] 

When {\bf K=B}, Eq. (\ref{wei3d}) is the PF of the 2D Ising model, and {\bf B} is (eg., Dixon et al., 2001),

\[{\bf B} = \left[ {\begin{array}{*{20}{c}}
{\bf A}&{\bf I}&0&0& \cdot & \cdot & \cdot &{\bf I}\\
{\bf I}&{\bf A}&{\bf I}&0& \cdot & \cdot & \cdot &0\\
0&{\bf I}&{\bf A}&{\bf I}& \cdot & \cdot & \cdot &{}\\
0&0&{\bf I}&{}&{}&{}&{}&{\bf I}\\
{\bf I}&0& \cdot & \cdot & \cdot &{}&{\bf I}&{\bf A}
\end{array}} \right]\]

with the {$(r,s)$-th} ($r=1,2,..., N$, $s=1,2,..., N$) eigenvalue is (eg., Berlin and Kac, 1952),

\[{E_{r,s}} = 2z\left\{ {\cos \left[ {\frac{{2\pi }}{N}(r - 1)} \right] + \cos \left[ {\frac{{2\pi }}{N}(s - 1)} \right]} \right\}\]

When {\bf K=C}, Eq. (\ref{wei3d}) is the PF of the 3D Ising model, and {\bf C} is (eg., Dixon et al., 2001),

\[\bf C = \left[ {\begin{array}{*{20}{c}}
{\bf B}&{\bf I}&{\bf O}& \cdots & \cdots &{}&{}&{\bf I}\\
{\bf I}&{\bf B}&{\bf I}&{}&{}&{}&{}&{\bf O}\\
{\bf O}&{\bf I}&{\bf B}&{}&{}&{}&{}&{}\\
{}&{}&{}&{}&{}&{}&{}&{\bf O}\\
{\bf O}&{}&{}&{}&{}&{}&{}&{\bf I}\\
{\bf I}&{\bf O}& \cdots & \cdots &{\bf O}&{}&{}&{\bf B}
\end{array}} \right]\]

with the {$(r,s,t)$-th} ($r=1,2,..., N$, $s=1,2,..., N$,$t=1,2,..., N$) eigenvalue is (eg., Berlin and Kac, 1952),

\[{E_{r,s,t}} = 2z\left\{ {\cos \left[ {\frac{{2\pi }}{N}(r - 1)} \right] + \cos \left[ {\frac{{2\pi }}{N}(s - 1)} \right] + \cos \left[ {\frac{{2\pi }}{N}(t - 1)} \right]} \right\}\]

Therefore, it is reasonable to infer from Eq. (\ref{wei3d}), {\bf A}, {\bf B}, {\bf C} and their eigenvalues that the form of $\frac{1}{N}\log Q_{_{3d}}(0,T)$ should be similar to that of $\frac{1}{N}\log Q_{_{1d}}(0,T)$ and $\frac{1}{N}\log Q_{_{2d}}(0,T)$.

\section{An EPF for the simple cubic Ising model}

From Eq. (\ref{Eq_infer}) it can be found that only $f(z)$ and $g(z)$ should be inferred to get an EPF for the simple cubic Ising model. In this section we will infer them through the connection between the lattice Green function (LGF) for the simple cubic lattice and that for the honeycomb lattice.  

Guttmann (2010) pointed out that there is a close relationship between the LGF of the $d$-dimensional hyper-cubic lattice and that of the ($d-1$)-dimensional diamond lattice. For $d=3$\footnote{Joyce (1994) had found that it is possible to express the LGF for 3D lattices in terms of the LGF for the 2D honeycomb lattice.}, one has,

\begin{equation}\label{guttman1}
{P_3}(z) = \frac{2}{\pi }\int_0^1 {\frac{{{Z_3}({t^2}{z^2}/{9})}}{{\sqrt {1 - {t^2}} }}} {\rm{d}}t
\end{equation}

where, 

\[
P_3(z) = \frac{1}{{{\pi ^3}}}\int_0^\pi  {\int_0^\pi  {\int_0^\pi  {\frac{{{\rm{d}}{\omega _1}{\rm{d}}{\omega _2}{\rm{d}}{\omega _3}}}{{1 - \frac{z}{3}\left( {\cos {\omega _1} + \cos {\omega _2} + \cos {\omega _3}} \right)}}} } }
\]

\[
\begin{array}{ll}
{{Z_3}({z^2}) = {P_{{\rm{honey}}}}(z)}&{ = \frac{1}{{{{(2\pi )}^2}}}\int_{ - \pi }^\pi  {\int_{ - \pi }^\pi  {\frac{{{\rm{d}}{\omega _1}{\rm{d}}{\omega _2}}}{{1 - \frac{{{z^2}}}{3}\left\{ {1 + \frac{2}{3}\left[ {\cos {\omega _1} + \cos {\omega _2} + \cos ({\omega _1} + {\omega _2})} \right]} \right\}}}} } }\\
{}&{ = \frac{1}{{{{(2\pi )}^2}}}\int_{ - \pi }^\pi  {\int_{ - \pi }^\pi  {\frac{{{\rm{d}}{\omega _1}{\rm{d}}{\omega _2}}}{{1 - \frac{{{z^2}}}{9}\left( {1 + 4{{\cos }^2}{\omega _1} + 4\cos {\omega _1}\cos {\omega _2}} \right)}}} } }\\
{}&{ = \frac{1}{{{{(2\pi )}^2}}}\int_0^{2\pi } {\int_0^{2\pi } {\frac{{{\rm{d}}{\omega _1}{\rm{d}}{\omega _2}}}{{1 - \frac{{{z^2}}}{9}\left( {1 + 4{{\cos }^2}{\omega _1} + 4\cos {\omega _1}\cos {\omega _2}} \right)}}} } }\\
{}&{ = \frac{1}{{{{(2\pi )}^2}}}\int_0^{2\pi } {\int_0^{2\pi } {\frac{{{\rm{d}}{\omega _1}{\rm{d}}{\omega _2}}}{{1 - \frac{{{z^2}}}{3}\left\{ {1 + \frac{2}{3}\left[ {\cos {\omega _1} + \cos {\omega _2} + \cos ({\omega _1} + {\omega _2})} \right]} \right\}}}} } }\end{array}
\]
where $P_3$, $P_{{\rm{honey}}}$ are the LGF of the simple cubic lattice and honeycomb lattice, respectively.

From Eq. (\ref{Eq_infer}) we can deduce the derivative $P(z)$ of $\frac{1}{N}\log Q_{_{3d}}(0,T)$ to $z$, 

\begin{equation}\label{guttman2}
P(z) \sim \frac{1}{{{\pi ^3}}}\int_0^\pi  {\int_0^\pi  {\int_0^\pi  {\frac{{{\rm{d}}{\omega _1}{\rm{d}}{\omega _2}{\rm{d}}{\omega _3}}}{{\frac{{f(z)}}{{g(z)}} - \left( {\cos {\omega _1} + \cos {\omega _2} + \cos {\omega _3}} \right)}}} } }
\end{equation}
where "$\sim$" means that we omit some items related to $f(z)$ and $g(z)$. 

According to Houtappel (1950), the PF for the honeycomb lattice Ising model is,
 
\begin{equation}\label{Houtappel}
\log {Q_{{\rm{honey}}}} = \frac{1}{{16{\pi ^2}}}\int_0^{2\pi } {\int_0^{2\pi } {\log \frac{1}{2}\left\{ {{{\cosh }^3}2z + 1 - {{\sinh }^2}2z\left[ {\cos {\omega _1} + \cos {\omega _2} + \cos ({\omega _1} + {\omega _2})} \right]} \right\}} } {\rm{d}}{\omega _1}{\rm{d}}{\omega _2}
\end{equation}

We can derive the derivative $P_{_{{\rm{honey}}}}$ of $\log {Q_{{\rm{honey}}}}$ to $z$ from Eq. (\ref{Houtappel}),

\begin{equation}\label{Houtappe2}
P_{_{{\rm{honey}}}}(z) \sim \frac{1}{{{{(2\pi )}^2}}}\int_0^{2\pi } {\int_0^{2\pi } {\frac{{{\rm{d}}{\omega _1}{\rm{d}}{\omega _2}}}{{\frac{{{{\cosh }^3}2z + 1}}{{{{\sinh }^2}2z}} - \left[ {\cos {\omega _1} + \cos {\omega _2} + \cos ({\omega _1} + {\omega _2})} \right]}}} }
\end{equation}

Combining Eq. (\ref{guttman1}), Eq. (\ref{Houtappe2}) and  Eq. (\ref{guttman2}), we can infer that,

\begin{equation}\label{wrq1}
\begin{array}{ll}
{f(z) \sim }&{\sqrt {2{{\cosh }^3}2z + 3{{\sinh }^2}2z + 2} }\\
{g(z) \sim }&{\sinh 2z}
\end{array}
\end{equation}

and the simplest form for $\frac{f(z)}{g(z)}$ is,

\begin{equation}\label{wrq2}
\frac{f(z)}{g(z)}\propto \frac{{\sqrt {2{{\cosh }^3}2z + 3{{\sinh }^2}2z + 2} }}{\sinh 2z}=\frac{{\sqrt {2{{\cosh }^3}2z + 3{{\sinh }^2}2z + 2} }}{\alpha\sinh 2z}
\end{equation}
where $\alpha$ is a constant. 

On the other hand, according to Guttmann (2010), $P_3(z)$ and $P_{_{{\rm{honey}}}}(z)$ can be expanded into series,

\begin{equation}\label{guttman3}
{P_{_{\rm{honey}}}}(z) = {\sum\limits_{n \ge 0} {{a_n}\left( {\frac{z}{3}} \right)} ^n}
\end{equation}

where, 
\[{a_n} = {\sum\limits_{j = 0}^n {\left( {\begin{array}{*{20}{c}}
n\\
j
\end{array}} \right)} ^2}\left( {\begin{array}{*{20}{c}}
{2j}\\
j
\end{array}} \right)\]

\begin{equation}\label{guttman4}
{P_3}(z) = {\sum\limits_{n \ge 0} {{b_n}\left( {\frac{z}{6}} \right)} ^n}
\end{equation}

where 
\[{b_n} = \left( {\begin{array}{*{20}{c}}
{2n}\\
n
\end{array}} \right){\sum\limits_{j = 0}^n {\left( {\begin{array}{*{20}{c}}
n\\
j
\end{array}} \right)} ^2}\left( {\begin{array}{*{20}{c}}
{2j}\\
j
\end{array}} \right)\]

From Eq. (\ref{guttman3}) and Eq. (\ref{guttman4}) we can infer $\alpha$ may be equal to $\sqrt{2}$, since there is a difference of 2 between their arguments. Hence,

 \begin{equation}\label{Eq_infer2}
 \begin{array}{ll}
\frac{1}{N}\log Q_{_{3d}}(0,T)=&\frac{1}{{2{\pi ^3}}}\int_0^\pi  \int_0^\pi  \int_0^\pi  \log [2\sqrt {2{{\cosh }^3}2z + 3{{\sinh }^2}2z + 2} \\
{}&{}\\
{}&\ \ \ \ \ - 2\sqrt{2}\sinh 2z \  (\cos\omega_1 + \cos \omega_2 + \cos\omega_3)]  {\rm{d}}\omega_1{\rm{d}}\omega_2{\rm{d}}\omega_3
\end{array}
\end{equation}

\begin{figure}[htb]
 \includegraphics[scale=0.5]{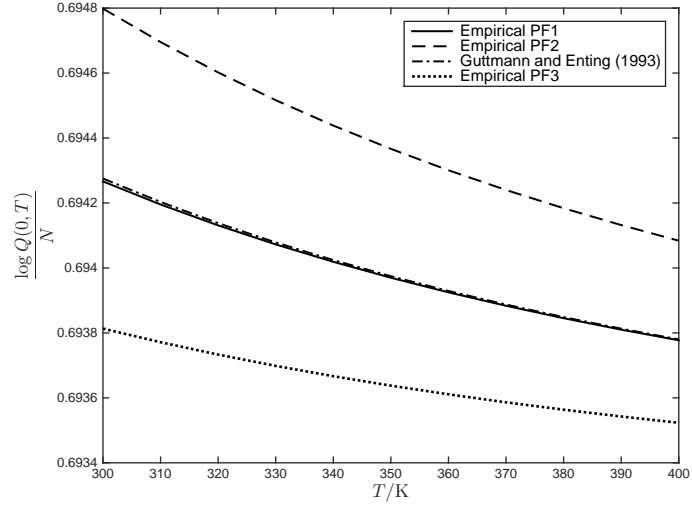}
 \caption{\footnotesize Comparison of the partition function (PF) calculated numerically from Eq. (\ref{Eq_infer2}) (Empirical PF1, $\alpha=\sqrt{2}$) with that from the high temperature expansions by Guttmann and Enting (1993). $\epsilon=1.0\times10^{-3}\mathrm{\  eV}$. Empirical PF2 is the PF from Eq. (\ref{wrq3}) (see details in subsection \ref{wrq_infer2}). Empirical PF3 is also from Eq. (\ref{Eq_infer2}) but $\alpha=1.67410$ (see details in subsection \ref{sub_dis_1}).}
\label{fig1}
\end{figure}

\begin{figure}[htb]
 \includegraphics[scale=0.5]{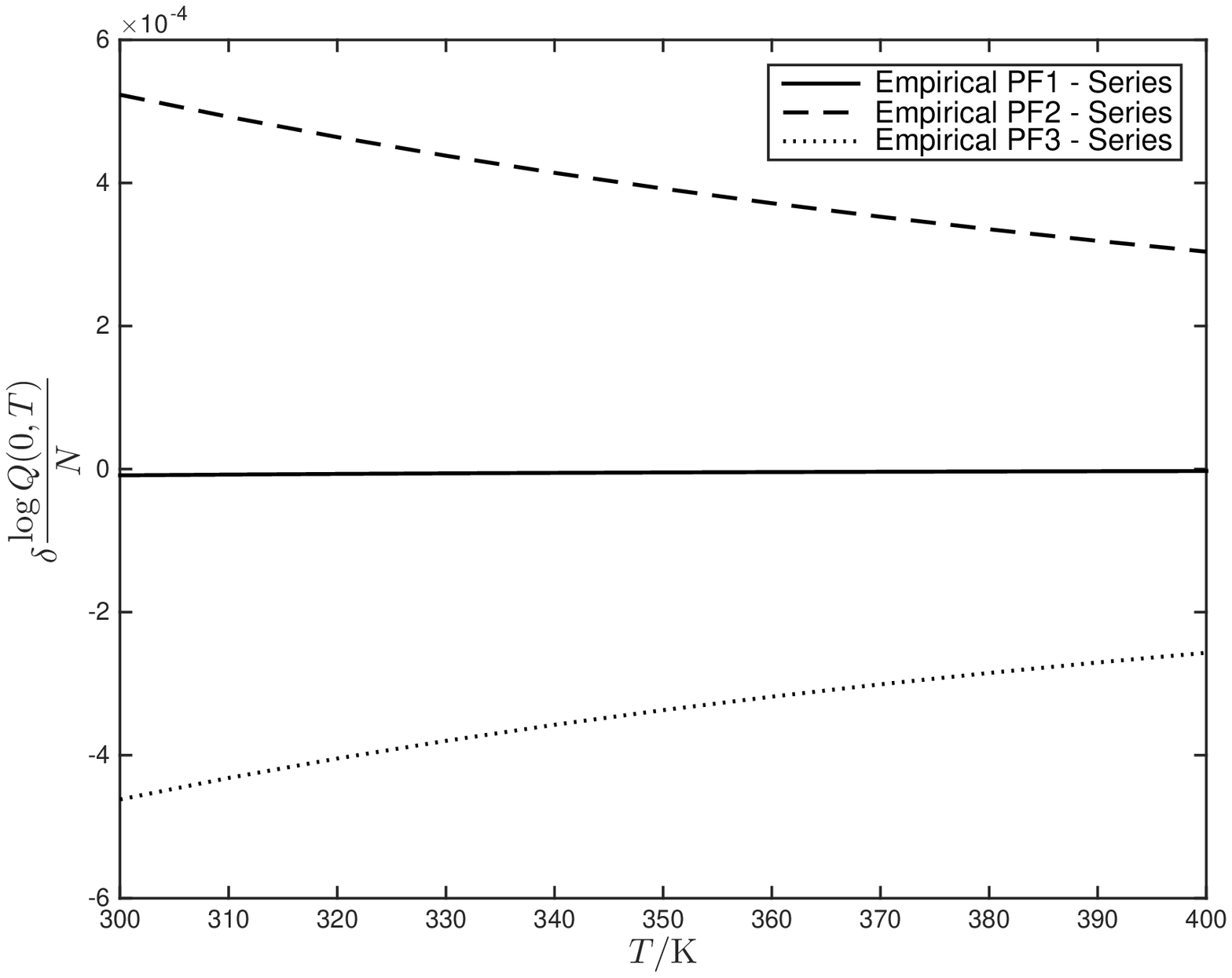}
 \caption{\footnotesize Differences between the PF that from the high temperature expansions (Series) and those from empirical PFs (Empirical PF1 is from Eq. (\ref{Eq_infer2}) with $\alpha=\sqrt{2}$); Empirical PF2 is from Eq. (\ref{wrq3}), Empirical PF3 is also from Eq. (\ref{Eq_infer2}) but $\alpha=1.67410$).  $\epsilon=1.0\times10^{-3}\mathrm{\  eV}$. }
\label{fig2}
\end{figure}

Fig. \ref{fig1} shows the comparison of the PF vs. temperature calculated from Eq. (\ref{Eq_infer2}) with that from the high temperature expansion by Guttmann and Enting (1993). It can be seen that these two PFs are consistent, which can be found clearly in Fig. \ref{fig2}. The consistence shows that the main properties of the simple cubic Ising model can be well understood by the EPF of Eq. (\ref{Eq_infer2}), at least in the high-temperature region. However, it can be found there is still a little difference between them, especially near the lower temperature end (see clearly in Fig. \ref{fig2}). We attribute mainly this to: (1) The errors from numerical integral method adopted to calculate the Eq. (\ref{Eq_infer2}); (2) The EPF itself needs improvement; (3) the series expanded may only hold at high temperatures.
 
\section{Discussion}
\subsection{The critical temperature $T_c$ from Eq. (\ref{Eq_infer2}) and the singularity}\label{sub_dis_1}

\ \ \ 

In this section, we discuss the critical temperature $T_c$ and the singularity of the simple cubic Ising model from Eq. (\ref{Eq_infer2}). For convenience, we rewrite Eq. (\ref{Eq_infer2}) as,

\begin{equation}\label{Eq_infer3}
\begin{array}{*{20}{r}}
\frac{1}{N}\log Q_{_{3d}}(0,T)=&\frac{1}{2}\log \left( {2\sqrt 2 \sinh 2z} \right) + \frac{1}{2{\pi ^3}}\int_0^\pi  \int_0^\pi  \int_0^\pi  \log [ {\frac{\sqrt {2{{\cosh }^3}2z + 3{{\sinh }^2}2z + 2} }{{\sqrt 2 \sinh 2z}}}  \\
{}&{}\\
{}&{ - {(\cos {\omega _1} + \cos {\omega _2} + \cos {\omega _3})} ]{\rm{d}}{\omega _1}{\rm{d}}{\omega _2}{\rm{d}}{\omega _3}}\\
{}&{}\\
{\sim}&\frac{1}{2{\pi ^3}}\int_0^\pi  \int_0^\pi  \int_0^\pi  \log [ {\frac{\sqrt {2{{\cosh }^3}2z + 3{{\sinh }^2}2z + 2} }{{\sqrt 2 \sinh 2z}}}  \\
{}&{}\\
{}& - {(\cos {\omega _1} + \cos {\omega _2} + \cos {\omega _3})} ]{\rm{d}}{\omega _1}{\rm{d}}{\omega _2}
\end{array}
\end{equation}
where "$\sim$" means the analytical $\frac{1}{2}\log \left( {2\sqrt 2 \sinh 2z} \right)$ is not taken into account.

Thus the internal energy $U$ is,

\begin{equation}\label{internal-energy}
\begin{array}{ll}
 U&=-\frac{\partial }{\partial\beta}[\frac{1}{N}\log Q_{_{3d}}(0,T)] \\
{}&{}\\
{}&\sim \frac{1}{{2{\pi ^3}}}\int_0^\pi  \int_0^\pi  \int_0^\pi  \frac{{{\rm{d}}{\omega _1}{\rm{d}}{\omega _2}{\rm{d}}{\omega _3}}}{\frac{{\sqrt {2{{\cosh }^3}2z + 3{{\sinh }^2}2z + 2} }}{{\sqrt 2 \sinh 2z}} -{(\cos {\omega _1} + \cos {\omega _2} + \cos {\omega _3})} } \\
{}&{}
\end{array}
\end{equation}
where "$\sim$" means the analytical parts, eg., the derivative of $\sqrt {2{{\cosh }^3}2z + 3{{\sinh }^2}2z + 2}/\sqrt 2 \sinh 2z$ to $\beta$, are not taken into account.

The integral of Eq. (\ref{internal-energy}) had been well studied in Joyce's series of articles. For example, in Joyce (1973), this integral is evaluated as,

 \begin{equation}\label{joyce-1973}
 \begin{array}{ll}
P(w) &=\frac{1}{{{\pi ^3}}}\int_0^\pi  \int_0^\pi  \int_0^\pi  \frac{{{\rm{d}}{\omega _1}{\rm{d}}{\omega _2}{\rm{d}}{\omega _3}}}{w -{(\cos {\omega _1} + \cos {\omega _2} + \cos {\omega _3})} }\\
{}&= \frac{{1 - 9{\xi ^4}}}{{{{(1 - \xi )}^3}(1 + 3\xi )}}{\left[ {\frac{2}{\pi }{\bf{K}}(k)} \right]^2}
\end{array}
\end{equation}
where ${\bf{K}}(k)$ is the complete elliptic integral of the first kind, and \[{k^2} = \frac{{16{\xi ^3}}}{{{{(1 - \xi )}^3}(1 + 3\xi )}}\]
with \[\xi  = {(1 + \sqrt {1 - {w^2}} )^{ - 1/2}}{(1 - \sqrt {1 - {w^2}/9} )^{1/2}}\]

For the sake of intuitiveness, we adopt the form of series solution for Eq. (\ref{internal-energy}) by Joyce (2001;2003), rather than Eq. (\ref{joyce-1973}), which reads,

\begin{equation}\label{internal-energy2}
\frac{1}{{\pi ^3}}\int_0^\pi  \int_0^\pi  \int_0^\pi  \frac{{{\rm{d}}{\omega _1}{\rm{d}}{\omega _2}{\rm{d}}{\omega _3}}}{w -{(\cos {\omega _1} + \cos {\omega _2} + \cos {\omega _3})} }=\sum\limits_{n = 0}^\infty  {{A_n}{{(w - 3)}^n}}  + {(w - 3)^{1/2}}\sum\limits_{n = 0}^\infty  {{B_n}} {(w - 3)^n}
\end{equation}
where $A_n$, $B_n$ are constants satisfy some recurrence relations (Joyce, 2001; 2003). This analytic continuation formula holds in the immediate neighborhood of the points $w =\pm 3$. 

If we let $A(w)=\sum\limits_{n = 0}^\infty  A_n(w - 3)^n$, $B(w)=\sum\limits_{n = 0}^\infty  B_n(w - 3)^n$, then

\begin{equation}\label{internal-energy2}
\frac{1}{{\pi ^3}}\int_0^\pi  \int_0^\pi  \int_0^\pi  \frac{{{\rm{d}}{\omega _1}{\rm{d}}{\omega _2}{\rm{d}}{\omega _3}}}{w -{(\cos {\omega _1} + \cos {\omega _2} + \cos {\omega _3})} }=A(w)  + (w - 3)^{1/2}B(w)
\end{equation}

And thus specific heat $C(0,T)$ is,

\begin{equation}\label{specific-heat}
C(0,T)=-k\beta^2\frac{\partial U}{\partial\beta}\sim \frac{1}{\sqrt{\frac{{\sqrt {2{{\cosh }^3}2z + 3{{\sinh }^2}2z + 2} }}{{\sqrt 2 \sinh 2z} }-3}}
\end{equation}
where "$\sim$" means that we omit the analytical parts, such as $\frac{\partial A(w)}{\partial \beta}$, and so on.

Obviously, the critical temperature $T_c$ is such that,

\[
\begin{array}{l}
\frac{{\sqrt {2{{\cosh }^3}2z + 3{{\sinh }^2}2z + 2} }}{{\sqrt 2 \sinh 2z} }-3=0\\
{}\\
\frac{\epsilon}{kT_c}=z=\cosh^{-1}[\frac{1}{4} (17-3\sqrt{17})]/2\approx 0.2772122674
\end{array}
\]

This $\frac{\epsilon}{kT_c} \approx 0.2772122674 > 0.221654626(5)$, the later is from a recent high-resolution
Monte Carlo study on a finite-size simple cubic Ising model (Xu et al., 2018). If we keep the form of Eq. (\ref{wrq2}) and let $\frac{\epsilon}{kT_c} = 0.221654626(5)$, $\alpha\approx 1.67410$. Fig. \ref{fig1} shows the corresponding PF vs. temperature (Empirical PF3) calculated. It can be seen that this PF is less than that from the high temperature expansions by Guttmann and Enting (1993), which can be found clearly in Fig. \ref{fig2}. If we let $\alpha=\sqrt{3}$ in  Eq. (\ref{wrq2}), the $\frac{\epsilon}{kT_c}=z=\cosh^{-1}[\frac{1}{2} (13-3\sqrt{13})]/2\approx 0.2124919895 < 0.221654626(5)$. Hence we infer that $\alpha\in(\sqrt{2},\sqrt{3})$ for the true EPF of the simple cubic Ising model.

It can also be found from Eq. (\ref{specific-heat}) that the specific heat approaches infinity as $|T-T_c|\rightarrow 0$. However, this is different from that in 2D case in which the specific heat approaches infinity logarithmically. 
  
\subsection{A pretty EPF in form}\label{wrq_infer2}

Eq. (\ref{Eq_infer2}) is not beautiful enough. We can rewrite the PF of the 1D and 2D Ising model as follows,

 \begin{equation}\label{KW2}
\begin{array}{lr}
 \frac{1}{N}\log Q_{_{1d}}(0,T) &=\frac{1}{2}\log (2\sinh z) + \frac{1}{{2\pi }}\int_0^\pi  \log (2\coth z\cosh z \\
  {}&- 2\cos {\omega _1}){\rm{d}}{\omega _1}
\end{array}
\end{equation}

 \begin{equation}\label{Onsager2}
\begin{array}{lr}
\frac{1}{N}\log Q_{_{2d}}(0,T)& = \frac{1}{2}\log (2\sinh 2z) + \frac{1}{{2{\pi ^2}}}\int_0^\pi  \int_0^\pi \log (2\cosh 2z\coth 2z \\
{}& - 2\cos {\omega _1} - 2\cos {\omega _2}){\rm{d}}{\omega _1}{\rm{d}}{\omega _2}  
\end{array}
\end{equation}

According to the section \ref{sec2}, the most natural representation of PF for the 3D Ising  model should be,

 \begin{equation}\label{wrq3}
\begin{array}{lr}
\frac{1}{N}\log Q_{_{3d}}(0,T)& = \frac{1}{2}\log (2\sinh 3z) + \frac{1}{{2{\pi ^3}}}\int_0^\pi \int_0^\pi \int_0^\pi  \log (2\cosh 3z\coth 3z \\
{}& - 2\cos {\omega _1} - 2\cos {\omega _2}-2\cos {\omega _3}){\rm{d}}{\omega _1}{\rm{d}}{\omega _2}  {\rm{d}}{\omega _3}
\end{array}
\end{equation}

 Fig. \ref{fig1} shows the corresponding PF vs. temperature calculated (Empirical PF2) . It can be seen that this PF is greater than that from the high temperature expansion by Guttmann and Enting (1993), which can be found clearly in Fig. \ref{fig2}.  This shows numerically Eq. (\ref{wrq3}) could not be the representative of 3D Ising  model at least in the high-temperature region, although it looks pretty and lovable.

 \subsection{Future work}
 
 As can be seen from the above, there is still a little difference between the EPF and that from the high temperature expansion by Guttmann and Enting (1993), especially near the lower temperature end. This implies that we can improve the EPF.  A possible way is that we can compare the exact results from the finite-size 3D Ising model, in which the $\alpha$ in the Eq. (\ref{wrq2}), or $f(z)$ and $g(z)$ in the Eq. (\ref{Eq_infer}) can be determined accurately. For example, an exact PF for the finite-size 3D Ising model can be from Wei (2018) with the improvement of computing power in the future. 
 
 \section{Conclusions}

It is still a challenge to obtain a practicable and exact PF for the 3D Ising model in a short time.  

We analyze: (1) the PF for the 1D and 2D model; (2) the general PF expression for the finite-size Ising model (Wei, 2018); (3) the eigenvalues of matrix in the Ising Hamiltonian; (4) the connection between the LGF for the simple cubic lattice and that for the honeycomb lattice. Based on the analysis above, we infer an EPF for the simple cubic Ising model in the absence of an external field. This EPF is consistent well numerically with the result from high temperature expansion, and the specific heat from this EPF approaches infinity non-logarithmically at the critical temperature, but the critical temperature is greater than those from numerical simulations. 

It is shown here that a reasonable EPF is helpful for understanding the properties of the 3D Ising model although it has no rigorous derivation.

\vspace{10em}

{\Large\bf  Acknowledges}

\vspace{2em}

We are grateful to Professor Perk, J. H. for his interest, advices, and comments.

\vspace{10em}


\ \ 

\end{document}